\documentclass[12pt]{article}

\usepackage{latexsym, amsmath, amssymb}
\usepackage{graphicx}
\usepackage{psfrag}
\usepackage[square, sort&compress, numbers, comma]{natbib}

\newcounter{subequation}[equation]
\makeatletter

\def\bcite{\@ifnextchar [{\@tempswatrue\@bcitex}{\@tempswafalse\@bcitex[]}}
\def\@bcitex[#1]#2{\if@filesw\immediate\write\@auxout{\string\citation{#2}}\fi
  \let\@bcitea\@empty
  \@bcite{\@for\@bciteb:=#2\do
    {\@bcitea\def\@bcitea{,\penalty\@m\ }%
     \def\@tempa##1##2\@nil{\edef\@bciteb{\if##1\space##2\else##1##2\fi}}%
     \expandafter\@tempa\@bciteb\@nil
     \@ifundefined{b@\@bciteb}{{\reset@font\bf ?}\@warning
       {Citation `\@bciteb' on page \thepage \space undefined}}%
     \hbox{\csname b@\@bciteb\endcsname}}}{#1}}
\def\@bcite#1#2{{#1\if@tempswa , #2\fi}}

\def\thesubequation{\theequation\@alph\c@subequation}
\def\@subeqnnum{{\rm (\thesubequation)}}
\def\slabel#1{\@bsphack\if@filesw {\let\thepage\relax
   \xdef\@gtempa{\write\@auxout{\string
      \newlabel{#1}{{\thesubequation}{\thepage}}}}}\@gtempa
   \if@nobreak \ifvmode\nobreak\fi\fi\fi\@esphack}
\def\subeqnarray{\stepcounter{equation}
\let\@currentlabel=\theequation\global\c@subequation\@ne
\global\@eqnswtrue
\global\@eqcnt\z@\tabskip\@centering\let\\=\@subeqncr
$$\halign to \displaywidth\bgroup\@eqnsel\hskip\@centering
  $\displaystyle\tabskip\z@{##}$&\global\@eqcnt\@ne
  \hskip 2\arraycolsep \hfil${##}$\hfil
  &\global\@eqcnt\tw@ \hskip 2\arraycolsep
  $\displaystyle\tabskip\z@{##}$\hfil
   \tabskip\@centering&\llap{##}\tabskip\z@\cr}
\def\endsubeqnarray{\@@subeqncr\egroup
                     $$\global\@ignoretrue}
\def\@subeqncr{{\ifnum0=`}\fi\@ifstar{\global\@eqpen\@M
    \@ysubeqncr}{\global\@eqpen\interdisplaylinepenalty \@ysubeqncr}}
\def\@ysubeqncr{\@ifnextchar [{\@xsubeqncr}{\@xsubeqncr[\z@]}}
\def\@xsubeqncr[#1]{\ifnum0=`{\fi}\@@subeqncr
   \noalign{\penalty\@eqpen\vskip\jot\vskip #1\relax}}
\def\@@subeqncr{\let\@tempa\relax
    \ifcase\@eqcnt \def\@tempa{& & &}\or \def\@tempa{& &}
      \else \def\@tempa{&}\fi
     \@tempa \if@eqnsw\@subeqnnum\refstepcounter{subequation}\fi
     \global\@eqnswtrue\global\@eqcnt\z@\cr}
\let\@ssubeqncr=\@subeqncr
\@namedef{subeqnarray*}{\def\@subeqncr{\nonumber\@ssubeqncr}\subeqnarray}
\@namedef{endsubeqnarray*}{\global\advance\c@equation\m@ne%
                           \nonumber\endsubeqnarray}

\renewcommand\maketitle{\par
  \begingroup
    \if@twocolumn
      \ifnum \col@number=\@ne
        \@maketitle
      \else
        \twocolumn[\@maketitle]%
      \fi
    \else
      \newpage
      \global\@topnum\z@   
      \@maketitle
    \fi
    \thispagestyle{plain}\@thanks
  \endgroup
  \setcounter{footnote}{0}%
  \global\let\thanks\relax
  \global\let\maketitle\relax
  \global\let\@maketitle\relax
  \global\let\@thanks\@empty
  \global\let\@author\@empty
  \global\let\@date\@empty
  \global\let\@title\@empty
  \global\let\title\relax
  \global\let\author\relax
  \global\let\date\relax
  \global\let\and\relax
}

\makeatother

\DeclareFontFamily{OT1}{rsfs11}{}
\DeclareFontShape{OT1}{rsfs11}{m}{n}{ <-> rsfs11 }{}
\DeclareMathAlphabet{\mathscript}{OT1}{rsfs11}{m}{n}

\numberwithin{equation}{section}

\newcommand{\gtlt}{\mathrel{\raise2.5pt\hbox{\oalign{$\scriptstyle>$\crcr
$\scriptstyle<$}}}}

\newcommand{\e}{{\mathrm e}}

\newcommand{\ie}{{\it i.e.}~}

\newcommand{\bo}{\raise-0.4mm\hbox{$\Box$}}              

\newcommand{\pt}{\partial}

\newcommand{\be}{\begin{equation}}
\newcommand{\ee}{\end{equation}}
\renewcommand{\[}{\begin{equation}}
\renewcommand{\]}{\end{equation}}

\newcommand{\nn}{\nonumber}
\newcommand{\bea}{\begin{eqnarray}}
\newcommand{\eea}{\end{eqnarray}}
\newcommand{\bsea}{\begin{subeqnarray}}
\newcommand{\esea}{\end{subeqnarray}}

\renewcommand{\tt}{\rightarrow} 

\def\a{\alpha}
\def\b{\beta}

\def\c{\chi}

\renewcommand{\d}{\mathrm{d}}

\def\e{\epsilon}

\def\f{\phi}

\def\m{\mu}
\def\n{\nu}

\def\r{\rho}
\def\s{\sigma}

\def\t{\tau}

\def\L{\Lambda}

\def\cL{{\cal L}}

\def\cT{{\cal T}}

\begin{document}

\begin{titlepage}
\begin{flushright}
DAMTP-2007-73
\end{flushright}
\vspace{.5cm}
\begin{center}
\baselineskip=16pt {\huge Bouncing Negative-Tension Branes
\\ }
\vspace{10mm}
{\large Jean-Luc Lehners and Neil Turok}
\vspace{15mm}

{\small\it  DAMTP, CMS, Wilberforce Road, CB3 0WA, Cambridge, UK.
} \\
\vspace{2cm}

\end{center}

\abstract{Braneworlds, understood here as double domain wall
spacetimes, can be described in terms of a linear harmonic function,
with kinks at the locations of the boundary branes. In a dynamical
setting, there is therefore the risk that the boundary brane of
negative tension, at whose location the value of the harmonic
function is always lowest, can encounter a zero of this harmonic
function, corresponding to the formation of a singularity. We show
that for certain types of brane-bound matter this singularity can be
avoided, and the negative-tension brane can shield the bulk
spacetime from the singularity by bouncing back smoothly before reaching
the singularity. In our analysis we compare the 5- and 4-dimensional
descriptions of this phenomenon in order to determine the validity
of the moduli space approximation. }

\vspace{2mm} \vfill \hrule width 2.3cm \vspace{2mm}{\footnotesize
\noindent \hspace{-9mm}
 E-mail: \texttt{j.lehners@damtp.cam.ac.uk, n.g.turok@damtp.cam.ac.uk.} }

\end{titlepage}

\setcounter{page}{2}

\section{Introduction}

Recently, a solution to the classical equations of motion of
heterotic M-theory was found, which describes a ``non-singular''
collision of the two boundary branes \cite{Lehners:2006pu}. By
non-singular we mean here that the volume of the internal
Calabi-Yau manifold, as well as the scale factors on the branes
remain finite and non-zero at the collision, with only the
orbifold dimension shrinking to a point. Since domain wall
solutions are usually described in terms of a linear harmonic
function, one might however expect on general grounds that in a
time-dependent context a zero of the harmonic function and thus a
spacetime singularity might be encountered at some other point in
the evolution. This is indeed the case. The zero of the harmonic
function in fact corresponds to a timelike naked singularity,
which the negative-tension brane runs into in the absence of
matter on the branes. This is the instability described by Gibbons
{\it et al.} \cite{Gibbons:2005rt} and by Chen {\it et al.}
\cite{Chen:2005jp}\footnote{It was shown in \cite{Lehners:2005su}
that static Ho\v{r}ava-Witten braneworlds are stable subject to
perturbations of finite energy. However, the time-dependent
configurations described in \cite{Chen:2005jp} and
\cite{Lehners:2006pu} differ from the static configuration by a
homogeneous, infinite-energy perturbation.}.

However, in the presence of a small amount of certain types of
brane-bound matter, the negative-tension brane bounces off the
naked singularity {\it without} touching it. This behaviour is
only possible due to the peculiar properties of gravity on a brane
of negative tension, and in a sense one can say that in these
cases the naked singularity acts repulsively with respect to the
negative-tension brane. Thus, and perhaps paradoxically, the
negative-tension boundary brane can have a stabilising effect by
shielding the bulk spacetime from the naked singularity that
corresponds to the zero of the harmonic function (note that
because the negative-tension brane corresponds to a trough-like
kink, it is always the negative-tension brane, rather than the
positive-tension one, which will be the closest to a zero of the
harmonic function). It was shown in \cite{Lehners:2006ir} that
from a 4d effective point of view, the bounce of the
negative-tension brane corresponds to a reflection of the solution
trajectory off a boundary of moduli space. This reflection has the
consequence of converting entropy perturbations into curvature
perturbations \cite{Lehners:2007ac}, and is thus rather
significant in the context of ekpyrotic \cite{Khoury:2001wf} or
cyclic \cite{Steinhardt:2001st} cosmological models. In the
present paper we study the conditions for such a bounce to occur
in greater generality. What we find is that a certain inequality,
involving the trace of the brane matter stress-energy tensor and
its coupling to the scalar supporting the domain walls, has to be
satisfied in order for a bounce to be possible.

We will study the conditions for a bounce both in 5 dimensions and
using the 4d moduli space approximation. In the study of
higher-dimensional braneworlds, it is often useful to resort to a
4d effective description, since higher-dimensional settings are
often quite far removed from one's intuition. It is therefore
crucial to determine the validity of the effective theory. We will
do this by comparing the description of the bounce of the
negative-tension brane from a 5-dimensional point of view with the
description of the same phenomenon in the 4-dimensional moduli
space approximation, in the presence of various types of
brane-bound matter.

\section{Domain Walls in 5 Dimensions}

We will consider scalar-gravity theories with an exponential
scalar potential. The action is given by \bea S &=& \int_{5d}
\sqrt{-g} \, [R - \frac{1}{2}(\pt \phi)^2 - 6\a^2(3\b^2-2)\, e^{2
\b\phi}] \nn
\\ [1ex] && + 12 \a \int_{4d, \, y=-1} \sqrt{-g}\, e^{\b\phi} -
12 \a \int_{4d, \, y=+1} \sqrt{-g}\, e^{\b \phi},
\label{Action5d} \eea where $\a$ is a positive constant that
can be adjusted by a shift in the scalar $\phi$ (we will choose
a convenient value later on) and $\beta$ determines the
self-coupling of $\phi.$ Theories of this type are
well-motivated in a supergravity context, where they can arise
after flux compactification {\it \`{a} la} Scherk-Schwarz, see
for example \cite{Stelle:1998xg}.
Typically, the domain wall action is given by a
worldvolume-weighted superpotential \be \mp \int_{4d, \,
y=\pm1} \sqrt{-g}\,W(\phi), \ee where here $W(\phi)=12\a e^{\b
\phi}.$ This superpotential is then related to the potential
$V(\phi)=6\a^2(3\b^2-2)\, e^{2 \b\phi}$ by the usual
supergravity relationship \be V=\frac{1}{8}[(\frac{\pt W}{\pt
\phi})^2-\frac{2}{3}W^2], \ee see \cite{Bergshoeff:2000zn} and
the appendix of \cite{Lehners:2007xa} for more details.
The case $\b = -1$ corresponds to heterotic M-theory in its
simplest consistent truncation \cite{HW1,HW2,LOSW1,LOSW2};
$e^{\phi}$ then parameterises the volume of the internal
Calabi-Yau manifold.

The static vacuum of the theory above is given by a domain wall
spacetime of the form \bea
\d s^2 &=& h^{2/(6\b^2 -1)}(y)\,\big[B^2 \,(-\d \t^2 + \d \vec{x}^2) + A^2 \,\d y^2\big], \nn \\
 e^{\f} &=& A^{-1/\b}\, h^{-6\b/(6\b^2-1)}(y), \label{domainwall} \nn \\
h(y) &=& \a\,(6\b^2-1) y + D, \eea where $A$, $B$ and $D$ are
arbitrary constants and $h(y)$ is a linear harmonic function.
The $y$ coordinate is taken to span the orbifold
$S^1/\mathbb{Z}_2$ with fixed points at $y=\pm 1$. In the
`upstairs' picture of the solution, obtained by
$\mathbb{Z}_2$-reflecting the solution across the branes, there
is a downward-pointing kink at $y=-1$ and an upward-pointing
kink at $y=+1$. These ensure the junction conditions are
satisfied, with the negative-tension brane being located at
$y=-1$ and the positive-tension brane at $y=+1$. The coordinate
system used above is only a good coordinate system when \be
\b^2
> \frac{1}{6}, \ee and we will restrict our analysis to this range of
$\b$ (as discussed recently in \cite{Palma:2007tu}, for certain physical
properties there are qualitative differences when $0 \leq \b^2 \leq \frac{1}{6}$).

\begin{figure}[t]
\begin{center}
\includegraphics{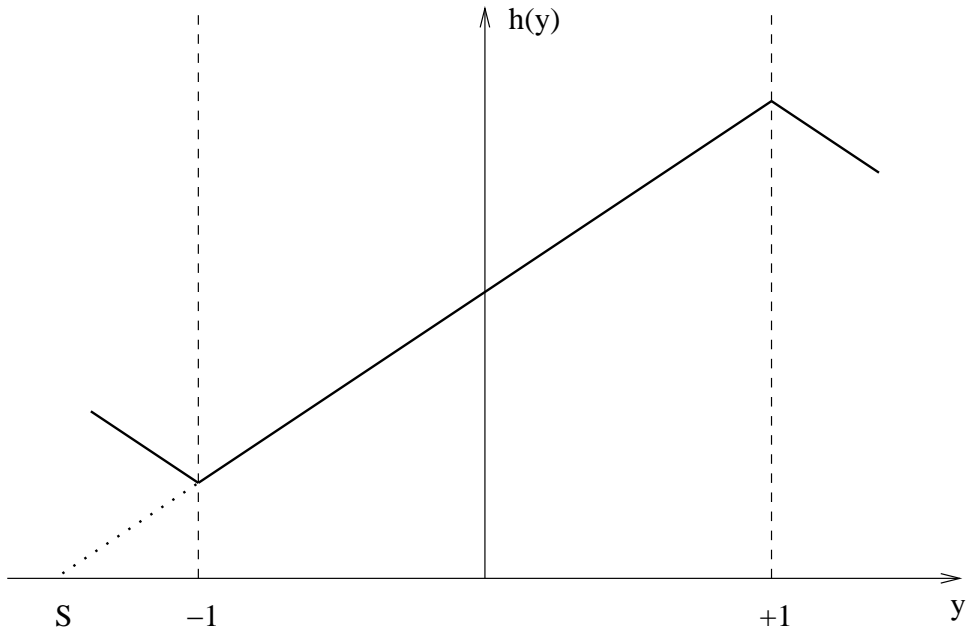} \caption{\label{fig1} {\small
The harmonic function $h(y)$, where $y$ is the coordinate on a
$S^1/\mathbb{Z}_2$ orbifold. In the absence of a negative-tension
brane at $y=-1,$ there would have been a singularity at $y=S.$}}
\end{center}
\end{figure}

The Ricci scalar is proportional to $h^{-12\b^2/(6\b^2-1)}$ and
thus the spacetime is singular at $h(y) = 0$. If we had only a positive-tension brane,
with a roof-type kink, this singularity would be at a finite
proper distance from the brane, and the spacetime would therefore
have a naked singularity. Usually, one avoids this problem by
cutting the spacetime off with a negative-tension brane placed in
between the positive-tension brane and the singularity, thereby
rendering the spacetime well-behaved, as we have already
anticipated by including two brane actions of opposite tension in
the action (\ref{Action5d}), see also Figure 1. In a
time-dependent context however, where the slope and the height of
the harmonic function can vary, there is still the risk that the
harmonic function can become zero at the location of the
negative-tension brane, thus causing a spacetime singularity to
form \cite{Gibbons:2005rt,Chen:2005jp}. In the next section we
will see that, in the presence of certain types of brane-bound
matter, this singularity can be avoided, with the negative-tension
brane bouncing back before it reaches the singularity.

\section{General Conditions for a Bounce of the Negative-Tension Brane}

In general, we add the following matter action at the location
of the negative-tension brane (at $y=-1$),
{\it i.e.} we add to equation (\ref{Action5d}) the term: \be +
\int_{4d,y= -1} \cL (g,\f,...), \ee where the dots represent
the matter contribution and we are allowing for a coupling to
the scalar $\f$. The junction conditions, which we are only
writing out here for the negative-tension brane, read (in this
section $' \equiv \frac{\pt}{\pt y}$ and $\dot{} \equiv
\frac{\pt}{\pt t}$) \bea a' &=& \a e^{n+\b\f} +\frac{1}{6} e^n
\cT^{0}_0
 \quad |_{y=-1} \label{jct1} \\ n' &=& \a
e^{n+\b\f} -\frac{1}{3} e^n \cT^{0}_0 +\frac{1}{6} e^n
\cT^{i}_i \quad |_{y=-1} \label{jct2}
\\ \f' &=& -6\a\b e^{n+\b\f} +\frac{1}{2} e^n \cT_{\f}
 \quad |_{y=-1}, \label{jct3} \eea where we have defined \bea \cT_{\mu \nu}
&\equiv& \frac{-1}{\sqrt{-g}} \frac{\delta \cL}{\delta g^{\mu \nu}} \\
 \cT_{\f} &\equiv& \frac{-1}{\sqrt{-g}} \frac{\delta
\cL}{\delta \f}, \eea with $\mu$ a brane worldvolume index.
Since the brane as well as the brane-bound matter are kept at
the fixed coordinate position $y=-1,$ we have $\cT_{\mu y}=0.$

We are only interested in whether or not the negative-tension
brane will bounce off the singularity, even if the bulk is
perturbed in the vicinity of this bounce. Therefore we will
choose a general metric and scalar field ansatz, which however
respects cosmological symmetry on the brane worldvolumes, so
that, on the branes, we have spatial homogeneity and isotropy:
\bea \d s^2
&=& e^{2n(t,y)} (-\d t^2 + \d y^2) + e^{2a(t,y)} \d\vec{x}^2 \\
e^{\f} &=& e^{\f(t,y)}. \eea With this metric ansatz we need
$\cT_{0i}=0$ for the $0i$ Einstein equation to be satisfied.

As a minimal requirement for a bounce to occur, there should be a
solution in which the negative-tension brane is momentarily
stationary ({\it i.e.} for which all first time derivatives are
zero at the location of the negative-tension brane), and in which
the second time derivative of the scale factor on the
negative-tension brane is positive. The $yy$ bulk Einstein
equation, which is an equation for the acceleration of the scale
factor $a$, is given by \be \label{MG55} 3\ddot{a} - 3
\dot{a}\dot{n} + \frac{1}{4} \dot{\phi}^2 + 6 \dot{a}^2 = 3a'^2 +
3 a' n' -\frac{1}{4}\phi'^2+ 9\a^2 (6\b^2-1)e^{2n+2\b\phi}. \ee We
can set first time derivatives to zero, since we are only
interested here in the moment of the bounce. Apart from the $ty$
Einstein equation (which is trivially satisfied at $y=-1$ since
every term involves a first time derivative), this equation is the
only one that involves only first derivatives with respect to $y$,
and so we can evaluate it at the location of the negative-tension
brane at the moment of the putative bounce by substituting in the
junction conditions (\ref{jct1})-(\ref{jct3}): \bea
3\ddot{a} &=& \frac{\a}{2} e^{2n+\b\f} (\cT^{\mu}_{\mu} + 3\b \cT_{\f}) \nn \\
&& -\frac{e^{2n}}{48} [4 (\cT^0_0)^2 - 4 \cT^0_0 \cT^i_i +3
(\cT_{\f})^2] \quad |_{y=-1;\textrm{bounce}} \label{B1} \eea The
first line is proportional to $\a$, and would therefore flip sign
on the positive-tension brane (where there would be additional
first time-derivative terms involved). The first line also
involves the trace of the matter stress-energy tensor. The second
line is proportional to the matter density squared, and can thus
be regarded as small compared to the first line. The second line
generally gives a negative contribution (it certainly does so when
the strong energy condition is satisfied).

If we want to have a bounce on the negative-tension brane, there
must be a positive contribution to $\ddot{a}$ from the first line
in (\ref{B1}), {\it i.e.} a necessary condition (but not
sufficient in general) is that \be \cT^{\mu}_{\mu} + 3\b \cT_{\f}
> 0. \label{B2} \ee This condition is not particularly difficult
to satisfy; we will give a few examples (and counter-examples) in
the next section. If equation (\ref{B2}) is satisfied, then one
also has to check that this contribution is dominant over the
second line in (\ref{B1}), which it is if the matter density is
sufficiently small. And one would of course have to extend the
solution to the rest of spacetime, which we simply assume here to
be feasible.

\section{Some Examples} \label{section5dexamples}

{\it Scalar Field}

\noindent Using the above equations, one can see that for a
scalar matter Lagrangian \be \cL = - \sqrt{-g} \frac{1}{2}
(\partial \s)^2 C(\phi), \ee where we allow for a coupling
$C(\f)$ and where we take $\s$ to depend only on time
(because of the assumed cosmological symmetry)
, we get a positive contribution to (\ref{B1}) when \be C - 3
\b C_{,\f} > 0. \ee Thus for a scalar field that doesn't couple
to $\f$, {\it i.e.} for which $C=1,$ we can expect a bounce;
however there will also be corrections to the geometry. Scalars
of this latter type are present in heterotic M-theory
\cite{LOSW2}. We will discuss the heterotic M-theory examples
in more detail in section \ref{sectionhetM}.

\vspace{0.7cm} \noindent {\it Gauge Field}

\noindent A vector gauge field localised on the brane is
represented by the Lagrangian \be \cL = - \sqrt{-g} C(\phi)
F_{\mu \nu}F^{\mu \nu}. \ee Here we assume the gauge field to
be abelian, and we use the usual electric-magnetic
decomposition \be F^{0i} = E_i \qquad F^{ij} = \e^{ijk} B_k.
\ee This leads to a stress-energy tensor
\bea \cT_{00} &=& -g_{00}(E^2 + B^2)C(\phi) \\
\cT_{0i} &=& -2 \e_{ijk} E^j B^k C(\phi) \\ \cT_{ij} &=&
[-2E_iE_j - 2B_iB_j + g_{ij} (E^2 +B^2)]C(\phi), \eea where we
have denoted $B=(B_i B^i)^{1/2}.$ We can immediately see that
the stress-energy tensor is traceless, \be \cT^{\m}_{\m} = 0.
\ee We also have \be \cT_\phi = C_{,\phi}(-2 E^2 +2 B^2). \ee
The $oi$ Einstein equation implies that $\cT_{0i},$ and thus
the Poynting vector, has to be zero. This will be the case if
we have an electric or a magnetic field only. Thus, from
(\ref{B2}), we can expect a bounce if \be \b C_{,\phi} < 0
\qquad \mathrm{and} \qquad B_i=0 \ee or if \be \b C_{,\phi}
> 0 \qquad \mathrm{and} \qquad E_i=0. \ee

On the other hand, it is easy to see that radiation alone, for
which the Poynting vector is zero on average, does not give
rise to a bounce, since then \footnote{In order to perform the
averaging, we are assuming here that $C_{,\phi}$ varies
slowly.} \be \langle E^2 \rangle = \langle B^2 \rangle. \ee In
that case the condition (\ref{B2}) cannot be fulfilled, as we
now have $\cT^{\mu}_{\mu} + 3 \b \cT_{\f} = 0$. However,
radiation also doesn't lead to a collapse; to first order in
the matter density it simply has no effect at all on whether we
have a bounce or not. It is only at second order in the energy
density that radiation contributes towards a collapse, as can
be seen from equation (\ref{B1}).

\vspace{0.7cm} \noindent {\it Perfect Fluid and Cosmological
Constant}

\noindent A perfect fluid with energy density $\rho$ can be described by the Lagrangian \be
\cL = - \sqrt{-g} \rho C(\f), \ee which leads to the stress-energy
tensor \cite{Fock} \bea \cT_{00} &=& -g_{00} \frac{1}{2} \rho C(\phi) \\
\cT_{ij} &=& g_{ij} \frac{1}{2} p C(\phi) \eea and \be \cT_{\phi}
= \rho C_{,\phi}, \ee where $p$ denotes the fluid's pressure. With an equation of state $p=w\rho$ and
$\rho>0$, we get a bounce if \be \b C_{,\phi} > \frac{1-3w}{6}C.
\ee

Note that due to the coupling to the scalar $\phi$,
radiation should not be represented as a perfect fluid with $w=\frac{1}{3}$, but rather
as a gauge field, as above. In fact, for that same reason, it is
doubtful to what extent the perfect fluid effective description is
accurate in general, except in the case of a cosmological
constant, which we write out explicitly here.

For a brane-localised cosmological constant $\L$, we would
consider \be \cL = - \sqrt{-g} 2 \L C(\f). \ee Then \be \cT_{\mu
\nu} = - \L g_{\mu \nu} C \ee and the condition (\ref{B2}) is
satisfied for \be  \L (\b C_{,\f} - \frac{2}{3} C) > 0. \ee  Thus,
for a positive cosmological constant $\L > 0$ we can expect a
bounce if the coupling is \be e^{c \f} \quad {\rm with} \quad \b c
> \frac{2}{3}. \ee If we have a negative cosmological constant, we can have a bounce
if the coupling is \be e^{c \f} \quad {\rm with} \quad \b c <
\frac{2}{3}. \ee Note that when $C=e^{\b \phi},$ the addition
of a cosmological constant corresponds to a de-tuning of the
brane tensions,
since it effectively changes the value of $\a$ in the brane
action at $y=-1$ in equation \ref{Action5d}.

\section{The Moduli Space Description}

For many reasons, not least because of our lack of intuition
about higher-dimensional settings and in order to make contact
with what we can observe at present, it is useful to have a
4-dimensional effective description of higher-dimensional
physics. An obvious question however is how much of the
higher-dimensional dynamics a 4d effective description can
capture. We will address this question by looking at the 4d
moduli space approximation for the examples presented in the
previous section.
The derivation of the moduli space action in this section will
be a generalisation to arbitrary $\b$ of the derivation in
\cite{Lehners:2006ir}, where it was performed for the case
$\b=-1.$

To implement the moduli space approximation, we simply promote
the moduli of the static solution (\ref{domainwall}) to
arbitrary functions of the brane conformal time $\t$, yielding
the ansatz:
\footnote{Note that the relationship between the coordinates
$(\t,x,y)$ used in this section and the coordinates $(t,x,y)$
used in the previous section is in general rather complicated.
We will not need the corresponding coordinate transformations
in this paper.}

\bea
\d s^2 &=& h^{2/(6\b^2 -1)}(\t,y)\,\big[B^2(\t) \,(-\d \t^2 + \d \vec{x}^2) + A^2(\t) \,\d y^2\big], \nn \\
 e^{\f} &=& A^{-1/\b}(\t)\, h^{-6\b/(6\b^2-1)}(\t,y), \label{domainwalltimemoduli} \nn \\
h(\t,y) &=& \a\,(6\b^2-1) y + D(\t), \qquad -1 \le y \le
+1.\eea

This ansatz satisfies the $\t y$ Einstein equation identically,
which is important, since otherwise the $\tau y$ equation would
act as a constraint \cite{Gray:2003vw}. Having defined the
time-dependent moduli, we would now like to derive the action
summarising their equations of motion. This is achieved by simply
plugging the ansatz (\ref{domainwalltimemoduli}) into the original
action (\ref{Action5d}), yielding the result (where we use the
notation $\dot{ } \equiv \pt/\pt \t$) \bea S_{\mathrm{mod}} &=& 6
\int_{4d} AB^2I_{\frac{3}{6\b^2-1}} \big[
\frac{1}{12\b^2}\Big(\frac{\dot{A}}{A}\Big)^2
-\Big(\frac{\dot{B}}{B}\Big)^2 -\frac{\dot{A}\dot{B}}{AB} \nn \\
&+& \frac{3\b^2-2}{(6\b^2-1)^2}
\frac{I_{\frac{-12\b^2+5}{6\b^2-1}}}{I_{\frac{3}{6\b^2-1}}}\,\dot{D}^2
-\frac{3}{6\b^2-1}\, \frac{I_{\frac{-6\b^2+4}{6\b^2-1}}
\dot{B}\dot{D}}{I_{\frac{3}{6\b^2-1}}B}\big],\qquad
\label{ActionMSA1} \eea where we have defined \be I_n =
\int_{-1}^{1} dy \ h^n =
\frac{1}{(n+1)\a(6\b^2-1)}[(D+\a(6\b^2-1))^{(n+1)}-(D-\a(6\b^2-1))^{(n+1)}].
\ee This action can be greatly simplified by introducing the
field redefinitions \bea
a_4^2 &\equiv& A\,B^2\,I_{\frac{3}{6\b^2-1}}, \\
e^{\sqrt{\frac{12\b^2}{3\b^2+1}}\psi} &\equiv& A\,(I_{\frac{3}{6\b^2-1}})^{3\b^2/(3\b^2+1)}, \\
(6\b^2-1) \chi &\equiv& - \int \d D\, \frac{
\big[(3\b^2-2)\,I_{\frac{-12\b^2+5}{6\b^2-1}}I_{\frac{3}{6\b^2-1}}
+\frac{9}{12\b^2+4}\,(I_{\frac{-6\b^2+4}{6\b^2-1}})^2\big]^{1/2}}{I_{\frac{3}{6\b^2-1}}}.
\label{DefinitionChi} \eea Note that $a_4$ has the
interpretation of being roughly the four-dimensional scale
factor, whereas $\psi$ and $\chi$ are four-dimensional scalars.
The definition (\ref{DefinitionChi}) can be rewritten as
stating that \be \sqrt{3\b^2+1}\,\d\chi = \frac{-\d
D}{(D+\a(6\b^2-1))^{(3\b^2-2)/(6\b^2-1)}\,(D-\a(6\b^2-1))^{(3\b^2-2)/(6\b^2-1)
}\, I_{\frac{3}{6\b^2-1}}}. \ee This expression can be
integrated to yield \be D = \a(6\b^2-1) \left[
\frac{(1+e^{2\sqrt{3\b^2+1}\chi})^{(6\b^2-1)/(3\b^2+1)}
+(1-e^{2\sqrt{3\b^2+1}\chi})^{(6\b^2-1)/(3\b^2+1)}}{(1+e^{2\sqrt{3\b^2+1}\chi})^{(6\b^2-1)/(3\b^2+1)}
-(1-e^{2\sqrt{3\b^2+1}\chi})^{(6\b^2-1)/(3\b^2+1)}}\right].
\label{relCmoduli} \ee In terms of $a_4$, $\psi$ and $\chi$ the
moduli space action (\ref{ActionMSA1}) then reduces to the
remarkably simple form \be \frac{1}{6} S_{\mathrm{mod}} =
\int_{4d} [-\dot{a_4}^2 + a_4^2 (\dot{\psi}^2 + \dot{\chi}^2)].
\label{ActionMSA2} \ee The minus sign in front of the kinetic
term for $a_4$ is characteristic of gravity, and in fact this
is the action for gravity with scale factor $a_4$ and two
minimally coupled scalar fields. Note that all the different 5d
theories, with different $\b$, are thus described by the same
4d effective theory to a first approximation. We will see
shortly however that the inclusion of brane-bound matter lifts
this degeneracy.

Useful expressions relating 4d and 5d quantities at the location of the
negative-tension brane are given by: \bea b_- &=&
(\a(6\b^2+2))^{1/(6\b^2+2)} \, a_4 \,
e^{-\sqrt{\frac{3\b^2}{3\b^2+1}}\psi}
 (-\sinh \sqrt{3\b^2+1}\c)^{1/(3\b^2+1)} \label{scalefactor} \\
e^{\f}_- &=& (\a(6\b^2+2))^{-6\b/(6\b^2+2)} \,
e^{-\frac{2}{\b}\sqrt{\frac{3\b^2}{3\b^2+1}}\psi} (-\sinh
\sqrt{3\b^2+1}\c)^{-6\b/(3\b^2+1)}, \eea where $b_-$ denotes
the brane scale factor $b_- = h^{1/(6\b^2-1)}(\t,y=-1)B(\t).$
Note that since $b_-$ is a positive quantity, the range of
$\chi$ should be restricted to $(-\infty,0].$ For simplicity we
will set $\a= 1/(6\b^2+2)$ in what follows; this can be done by a shift in $\phi.$
Also, in this section we always assume the coupling function
$C(\f)$ to be of the form \be C(\f) = e^{c \f}. \ee
In heterotic M-theory ($\b=-1$), where the volume of the
Calabi-Yau manifold is given by $e^{\phi},$ this corresponds to
the brane-bound matter fields coupling to a power of the volume
of the internal manifold.

Before continuing, let us present a brief argument which
partially explains the simplicity of the moduli space action
(\ref{ActionMSA2}). This arguments rests on the observation
that the original 5d action (\ref{Action5d}) is invariant under
the global scaling symmetry \bea g_{mn} &\tt& e^{2\e} g_{mn} \\
\phi &\tt& \phi - \frac{1}{\b}\e ,\eea where $\e$ is a constant
parameter. Under this symmetry, the moduli of the domain wall
solution (\ref{domainwalltimemoduli}) transform as \bea A &\tt&
e^\e A \\ B &\tt& e^\e B \\ D &\tt& D. \eea This in turn
corresponds to the transformations \bea a_4 &\tt& e^{3\e/2} a_4
\\ \psi &\tt& \psi + \sqrt{\frac{3\b^2+1}{12\b^2}}\e \\ \chi &\tt& \chi. \eea
Thus we see that this symmetry induces the shift symmetry in
$\psi.$ It is also interesting to note that the absence of an
implied shift symmetry in $\chi$ is consistent with the fact
that the range of $\chi$ is actually limited, as noted above,
and that the absolute value of $\chi$ is a meaningful quantity.

\vspace{0.7cm} \noindent {\it Scalar Field}

\noindent For a scalar field $\s$ coupling to the scalar $\f$
via $e^{c\f},$ with $c$ an arbitrary real number, we get an
addition to the
effective theory (\ref{ActionMSA2}) of \bea && -\sqrt{-g} e^{c\f} g^{00} \dot{\s}^2 \quad |_{y=-1} \\
&=& a_4^2 e^{-2(c/\b+1)\sqrt{\frac{3\b^2}{3\b^2+1}}\psi}
(-\sinh \sqrt{3\b^2+1}\c)^{(-6\b c + 2)/(3\b^2+1)} \dot{\s}^2.
\eea The equation of motion for $\s$ can be solved immediately
to give \be \dot\s = \frac{\s_0}{a_4^{2}}
e^{2(c/\b+1)\sqrt{\frac{3\b^2}{3\b^2+1}}\psi} (-\sinh
\sqrt{3\b^2+1}\c)^{(6\b c - 2)/(3\b^2+1)}, \label{sigma}\ee
where $\s_0$ is a constant. Also, the equation of motion \be
\frac{\ddot{a_4}}{a_4} = -\dot\psi^2 - \dot\chi^2 -
\frac{\s_0^2}{a_4^{4}}
e^{2(c/\b+1)\sqrt{\frac{3\b^2}{3\b^2+1}}\psi} (-\sinh
\sqrt{3\b^2+1}\c)^{(6\b c - 2)/(3\b^2+1)} \ee
together with the constraint\footnote{This constraint arises
from the time reparameterisation invariance of the action or,
equivalently, from the $00$ Einstein equation.} (Friedmann
equation)
\be \frac{\dot{a_4}^2}{a_4^2} = \dot\psi^2 +
\dot\chi^2 + \frac{\s_0^2}{a_4^{4}}
e^{2(c/\b+1)\sqrt{\frac{3\b^2}{3\b^2+1}}\psi} (-\sinh
\sqrt{3\b^2+1}\c)^{(6\b c - 2)/(3\b^2+1)} \ee lead to \be
a_4=\t^{1/2}. \ee If we then define a new time variable \be T
\equiv \ln{\t}, \ee the remaining equations of motion can be
expressed as \bea \psi_{,TT} + \frac{\s_0^2}{2} V_{,\psi} &=& 0 \\
\chi_{,TT} + \frac{\s_0^2}{2} V_{,\chi} &=& 0, \eea
or, equivalently, by the action \be \int_{4d} \psi_{,T}^2 +
\chi_{,T}^2 - \s_0^2 V(\psi,\chi). \ee
The effective potential is given by \be V =
e^{2(c/\b+1)\sqrt{\frac{3\b^2}{3\b^2+1}}\psi} (-\sinh
\sqrt{3\b^2+1}\c)^{(6\b c - 2)/(3\b^2+1)}. \label{effpotscalar}
\ee Therefore, as $\c \tt 0$ the effective potential blows up
and becomes repulsive if \be \b c < 1/3.
\label{MSAconditionscalar}\ee Thus the solution trajectory
effectively gets reflected off the $\chi=0$ plane which means
that the scale factor on the negative-tension brane starts
increasing again (see equation (\ref{scalefactor})), \ie the
negative-tension brane bounces. Condition
(\ref{MSAconditionscalar}) is the same as that derived above
from the 5d point of view in section \ref{section5dexamples}.

\vspace{0.7cm} \noindent {\it Gauge Field}

\noindent By adding a vector gauge field with Lagrangian \be
\cL = - \sqrt{-g} e^{c\phi} F_{\mu \nu}F^{\mu \nu} \quad
|_{y=-1}, \ee we obtain an effective theory described by the
action \be S = \int_{4d} [-\dot{a_4}^2 +
  a_4^2 (\dot{\psi}^2 + \dot{\chi}^2)
  -a_4^4 e^{-2(c/\b)\sqrt{\frac{3\b^2}{3\b^2+1}}\psi}(-\sinh \sqrt{3\b^2+1}\chi)^{-6\b c/(3\b^2+1)}F_{\mu \nu}F^{\mu \nu} ]. \ee Then we have the constraint \be
\frac{\dot{a_4}^2}{a_4^2} = \dot\psi^2 + \dot\chi^2 +a_4^2
e^{-2(c/\b)\sqrt{\frac{3\b^2}{3\b^2+1}}\psi}(-\sinh
\sqrt{3\b^2+1}\chi)^{-6\b c/(3\b^2+1)}(2E^2 + 2B^2) \label{G1} \ee
together with the equations of motion \bea && \hspace{-1.5cm}
\frac{\ddot{a_4}}{a_4} = -\dot{\psi}^2 - \dot{\chi}^2 \label{G2}\\
&& \hspace{-1.5cm} \ddot\psi + 2 \frac{\dot{a_4}}{a_4}\dot\psi +
\frac{1}{a_4^2}\frac{\pt}{\pt
  \psi}[e^{-2(c/\b)\sqrt{\frac{3\b^2}{3\b^2+1}}\psi}(-\sinh \sqrt{3\b^2+1}\chi)^{-6\b c/(3\b^2+1)}](-E^2 + B^2)=0 \label{G3} \\ &&
\hspace{-1.5cm} \ddot\chi + 2 \frac{\dot{a_4}}{a_4}\dot\chi +
\frac{1}{a_4^2}\frac{\pt}{\pt
  \chi}[e^{-2(c/\b)\sqrt{\frac{3\b^2}{3\b^2+1}}\psi}(-\sinh \sqrt{3\b^2+1}\chi)^{-6\b c/(3\b^2+1)}](-E^2 + B^2)=0 \label{G4} \\ &&
\hspace{-1.5cm} \pt^\mu [F_{\mu \nu}
e^{-2(c/\b)\sqrt{\frac{3\b^2}{3\b^2+1}}\psi}(-\sinh
\sqrt{3\b^2+1}\chi)^{-6\b c/(3\b^2+1)}]=0. \label{G5} \eea The
last equation, supplemented by the Bianchi identity \be \e^{\m \n
\r \s}\pt_{\n} F_{\r \s} = 0, \ee leads to \bea E &=& E_0
e^{2(c/\b)\sqrt{\frac{3\b^2}{3\b^2+1}}\psi}(-\sinh
\sqrt{3\b^2+1}\chi)^{6\b c/(3\b^2+1)}
\\ B_i &=& B_{i,0} \eea where $E_0$ and $B_{i,0}$ are constants.
The equations of motion for $\psi$ and $\chi$ can then be
rewritten as \bea \ddot\psi
+2\frac{\dot{a_4}}{a_4}\dot\psi + \frac{1}{a_4^2} V_{,\psi} &=& 0 \\
\ddot\chi + 2\frac{\dot{a_4}}{a_4}\dot\chi + \frac{1}{a_4^2}
V_{,\chi} &=& 0 \eea with the effective potential \bea V &=& E_0^2
e^{2(c/\b)\sqrt{\frac{3\b^2}{3\b^2+1}}\psi}(-\sinh
\sqrt{3\b^2+1}\chi)^{6\b c/(3\b^2+1)} \nn \\ &+& B_0^2
e^{-2(c/\b)\sqrt{\frac{3\b^2}{3\b^2+1}}\psi}(-\sinh
\sqrt{3\b^2+1}\chi)^{-6\b c/(3\b^2+1)}. \label{effpotgauge} \eea
Thus we can see that near $\c=0$ the effective potential blows up
and leads to a bounce of the negative-tension brane if we either
have an electric field and \be \b c < 0, \ee or if we have a
magnetic field and \be \b c
> 0. \ee This is in agreement with the 5d description of section
\ref{section5dexamples}. Also, if we consider radiation, for
which \be \langle E^2 \rangle = \langle B^2 \rangle, \ee it is
immediately apparent from equations (\ref{G2})-(\ref{G4}) that
it does not lead to a bounce. This is
again consistent with the 5d results derived earlier.

\vspace{0.7cm} \noindent {\it Cosmological Constant}

\noindent We can repeat the above analysis in the case of a
brane-localised cosmological constant $\Lambda,$ also coupling to
the scalar $\f$. In that case the effective action receives an
additional contribution of \bea && -\sqrt{-g} e^{c\f} 2\Lambda \quad |_{y=-1} \\
&=& -a_4^4 e^{-2(c/\b+2)\sqrt{\frac{3\b^2}{3\b^2+1}}\psi} (-\sinh
\sqrt{3\b^2+1}\c)^{(-6\b c+4)/(3\b^2+1)} 2\Lambda. \eea Therefore,
the effective potential is \be V = \L
e^{-2(c/\b+2)\sqrt{\frac{3\b^2}{3\b^2+1}}\psi} (-\sinh
\sqrt{3\b^2+1}\c)^{(-6\b c+4)/(3\b^2+1)} \label{effpotcosconstant}
 \ee and as $\c \tt
0,$ we get a bounce if \be \b c > 2/3 \quad (\Lambda > 0) . \ee
This is exactly the same requirement as that obtained from the 5d
point of view for a positive cosmological constant.

However, the case of a negative cosmological constant cannot be
reproduced within the 4d effective theory, as the effective
potential is negative in that case.

\section{Heterotic M-Theory Examples} \label{sectionhetM}

Heterotic M-theory corresponds to the special case $\b=-1,$ with
the scalar $\phi$ parameterising the volume of the internal
Calabi-Yau manifold \cite{LOSW1}. It is in this theory that the
colliding branes solution \cite{Lehners:2006pu}, which was briefly
discussed in the introduction and which motivated the present
work, was derived. The solution was described in a coordinate
system in which the bulk is static and the branes are moving. The
boundary conditions used correspond to requiring the brane scale
factors and the Calabi-Yau volume to be non-zero and finite at the
collision of the branes. This turns out to be equivalent to
imposing the relationship \cite{Lehners:2006pu} \be \f = 6 a. \ee
This condition relates the volume of the Calabi-Yau to the brane
scale factors, while reducing the number of independent fields to
two. This last feature enables one to derive a Birkhoff-like
theorem \footnote{For the case of general $\beta,$ a similar Birkhoff-like theorem can be derived if one imposes $\phi = -6 \beta a.$ The discussion in the present section can be generalised in a straightforward, but unilluminating way to having arbitrary $\beta.$}, which determines the bulk metric to be given by a
one-parameter time-independent family of metrics (the parameter
being the relative rapidity of the branes at the collision), with
the branes moving in this background geometry according to their
junction conditions. It is easy to see from the junction
conditions (\ref{jct1})-(\ref{jct3}) that we can keep the
requirement that $\f=6a,$ and thus the Birkhoff-like theorem
mentioned above, only if \be \cT^0_0 = \frac{1}{2} \cT_{\f}.
\label{BulkUnaltered} \ee Thus we can see that in general a very
specific coupling $C(\f)$ to the Calabi-Yau volume scalar is
required if we want the bulk spacetime to remain unaltered by the
presence of brane-bound matter (the brane trajectories will of
course be modified in any case).

For a brane-bound scalar, it is straightforward to see that the
bulk geometry is unaltered only if the coupling is \be C = e^\f.
\ee As shown in section \ref{section5dexamples}, there will also
be a bounce in this case, and the entire evolution can be
described exactly, since the bulk spacetime is given by the
solution described in \cite{Lehners:2006pu}. From the moduli space
point of view, we can note that the effective potential
(\ref{effpotscalar}) is independent of $\psi$ only for $C=e^\f,$
which coincides with the condition for the bulk geometry to be
unaltered. This can be understood by the fact that, if the
effective potential is independent of $\psi$, the scalar field
space trajectory reflects off the effective potential with the
same final angle as the incident angle, in a smoothed-out version
of a ``brick wall'' reflection at $\c=0,$ and therefore the
background trajectory is unchanged except for this symmetric
rounding off of the trajectory near the bounce of the
negative-tension brane. Thus, for scalar field matter, the 4d and
5d points of view are in perfect agreement. This can be traced
back to the fact that we are simply extending the moduli space by
one dimension, by adding an extra kinetic term, and therefore the
moduli space description should remain a good approximation.

Note that the scalars arising from the dimensional reduction of
the $E_8$ gauge fields in heterotic M-theory do not couple to the
Calabi-Yau volume, \ie they have $C=1$ \cite{LOSW2}. Scalars of
this type also make the negative-tension brane bounce. However,
the bulk geometry will be altered in this case, which is why it
might be of interest to calculate the resulting deformed geometry.

For gauge fields, condition (\ref{BulkUnaltered}) shows that the
bulk is unaltered only if \be -(E^2 + B^2)C = (-E^2 +
B^2)C_{,\phi}. \ee This can be satisfied either if we have an
electric field only ($ B=0$) with the coupling \be C=e^{\phi} \ee
or if we only have a magnetic field ($E=0$) and the coupling \be
C=e^{-\phi}. \ee  However, in both cases, the effective potential
(\ref{effpotgauge}) in the moduli space description is independent
of $\psi$ only if $C=1.$ While the moduli space approximation
correctly predicts whether or not a bounce occurs, the detailed
trajectory followed in this description is not perfectly symmetric
about the bounce (when the coupling is such that the bulk remains
unaltered), and hence not a perfect rendition of the 5d solution.

In fact, the $E_8$ gauge fields in heterotic M-theory couple with
$C=e^{\phi}$ \cite{LOSW2}. Their electric component therefore
contributes to a bounce, while also leaving the bulk geometry
unaltered, while their magnetic component rather contributes to a
crunch (and a deformation of the bulk geometry).

Again by inspection of (\ref{BulkUnaltered}), it is easy to see
that a brane-bound cosmological constant does not perturb the bulk
geometry if its coupling is given by $C=e^{-\phi}.$ In this case,
we simply have a de-tuning of the brane tension. This de-tuning
leads to a bounce if the cosmological constant is positive,
whereas it leads to a crunch if it is negative. Note that the
moduli space description yields a potential
(\ref{effpotcosconstant}) that is independent of $\psi$ only when
$C=e^{2\f},$ which is in disagreement with the 5d description.

\section{Conclusions}

In a dynamical braneworld setting, the negative-tension boundary
brane can encounter a zero of the harmonic function corresponding
to the formation of a singularity. However, we have shown that
this catastrophic encounter is avoided in the presence of a broad
range of brane-bound matter types and couplings to the
 scalar field supporting the domain walls, which make the
negative-tension brane bounce off the naked
singularity\footnote{Thus, we could say that we have a bang if no
observer is there to hear it, but no sound in the presence of the
right kind of observer!}. This leads us to the rather surprising
conclusion that negative-tension branes can stabilise braneworlds.

We have analysed the bounce of the negative-tension brane from
two points of view: firstly, we have looked at the 5d equations
of motion and junction conditions in the vicinity of the
bounce. And secondly, we have analysed the analogous situation
using the moduli space approximation. For scalar fields, the
two descriptions are in perfect agreement. This is because
adding a kinetic term is perfectly suited to the spirit of the
moduli space approximation. For gauge fields and for a positive
cosmological constant, the moduli space approach correctly
reproduces the 5d results for the bounce. However, when the
conditions are fulfilled for the 5d bulk to remain unaltered
and we hence know that the 4d trajectory should be perfectly
symmetric about the bounce, the 4d effective theory does not
reproduce this behaviour. And in the case of a negative
cosmological constant, the moduli space approach completely
disagrees with the 5d results. It seems clear that in case of a
disagreement, we should rather trust the 5d results.
In fact, our results indicate that in the case of a brane-bound
gauge field or a cosmological constant, the approximations used
in deriving the moduli space action are not really valid. In
these cases, there are non-flat directions in configuration
space which are easily accessible to the system under study,
and which are not described by the moduli space approximation.
Thus, even though the moduli space description can give
qualitatively correct results in describing the effects of a
gauge field or a positive cosmological constant, the detailed
quantitative analysis can be rather misleading, and one should
revert to a 5d description.

The types of brane-bound matter that are naturally present in
heterotic M-theory are scalar fields that do not couple to the
Calabi-Yau volume, and gauge fields with an $e^\f$ coupling. What
we found is that for this specific coupling, electric fields
contribute towards a bounce, while radiation has no effect and
magnetic fields rather contribute to a crunch. The scalars
contribute towards a bounce, and probably represent the best
candidates for stabilising the heterotic M-theory braneworld.

Finally, we would like to point out that it seems likely that
additional brane-bound matter will be produced by quantum effects
at the bounce of the negative-tension brane, and it would be
interesting to determine the properties of these new
contributions.

\section*{Acknowledgements}

The authors would like to thank Gary Gibbons, Paul McFadden,
Paul Steinhardt and Kelly Stelle for useful discussions. The authors are supported
by PPARC and the Centre for Theoretical Cosmology in Cambridge.

\bibliographystyle{apsrev}
\bibliography{BouncingBranes}

\end{document}